\documentclass{amsart}

\usepackage{graphicx}
\usepackage{amsmath}
\usepackage{amssymb}
\usepackage{amsfonts}
\usepackage{amscd}
\usepackage{amsthm}
\theoremstyle{plain}

\newtheorem{thm}{Theorem}
\newtheorem{prop}[thm]{Proposition}

\theoremstyle{definition}

\newtheorem{conj}[thm]{Conjecture}

\def\rank{\mathop{rank}}

\title{Asynchronous logic circuits and sheaf obstructions}
\author{Michael Robinson}

\begin{document}

\thanks{This work was supported under AFOSR FA9550-09-1-0643}

\begin{abstract}
This article exhibits a particular encoding of logic circuits into a
sheaf formalism.  The central result of this article is that there
exists strictly more information available to a circuit designer in
this setting than exists in static truth tables, but less than exists
in event-level simulation.  This information is related to the timing
behavior of the logic circuits, and thereby provides a ``bridge''
between static logic analysis and detailed simulation.
\end{abstract}

\maketitle

\section{Introduction}

Verification of asynchronous logic circuits usually requires extensive
simulation and appropriate test coverage.  This article presents a
novel technique for detecting certain behavioral properties of a logic
circuit using a less exhaustive structural analysis.  In this
analysis, wire delays are unknown and finite, but unlike the work of
others in this situation, the wire delays are implicit.  We do not
need to assume that they have a fixed value over time, and we never
specify them even as variables.  We show how potentially hazardous
race conditions (which often cause glitches or unwanted latching)
correspond to nontrivial first cohomology classes of a particular
sheaf that encodes this implicit timing model of the circuit.

\subsection{Historical discussion}

The synthesis of asynchronous logic circuits is an old subject, having
been studied in the earliest days of computing by Huffman
\cite{Huffman_1954}.  Although the benefits of using asynchronous over
synchronous hardware are substantial (better composibility of modules,
lower power usage, lower electromagnetic interference, faster speed),
its challenges have generally precluded its widespread acceptance.
Most of the difficulty of asynchronous design involves careful control
of delays within the circuit, and the avoidance of race conditions
\cite{Manohar_1995}.  However, even correct timing is insufficient to
ensure correct operation, due to subtleties involving switching
thresholds \cite{Berkel_1992}.  That said, an increasing number of
organizations use designs that incorporate some asynchronous portions
\cite{Edwards_2004}.  A few processors, for instance the ILLIAC
\cite{Robertson}, the Caltech Asynchronous Microprocessor
\cite{Martin_1989}, and the ACT11 \cite{Epson_2005}
have been constructed without global clocks.

The challenges of asynchronous design revolve around timing
instabilities and sensitivities, which usually mean that verification
requires exhaustive simulation.  As a result of both the benefits and
the challenges, a lively literature has grown up around the design and
verification of asynchronous logic.  There are essentially three major
threads of inquiry:
\begin{enumerate}
\item specification of a semantic or behavioral model of the circuit,
\item synthesis of the gate-level circuitry from this specification, and 
\item simulation of the circuit to verify its correct performance.
\end{enumerate}

Semantic specifications are often given using process algebras like
the $\pi$-calculus \cite{Parrow} or CSP \cite{Hoare_2004}.  The latter
gained some traction when Martin \cite{Martin_1986} described how to
compile a version of CSP into a gate-level logic circuit.  He later
refined this compilation to generate quasi-delay insensitive circuits
only \cite{Martin_1990}, a class of circuits that was later shown to
be Turing-complete \cite{Manohar_1995}.  Quasi-delay insensitivity
requires that the circuit be insensitive to wire delays, except for
certain pairs of signals that are assumed to arrive nearly
simultaneously.  It would seem that complete delay insensitivity would
be more desirable, but Brzozowksi \cite{Brzozowksi_1992} showed that
this unduly limits the circuits that can be constructed.  After these
initial efforts, the underlying theory of specifications for
asynchronous circuits has continued to develop, often drawing upon
methods in formal proof for mathematics \cite{Harrison_2008}, logic
\cite{Vasyukevich_2009}, and computer science \cite{Dalrymple_2008}.

In addition to Martin's work, other researchers have pursued
asynchronous logic synthesis approaches.  A detailed survey of some of
them is given in \cite{Davis_1997}.  Readers familiar with Petri nets
will find the unified synthesis treatment in \cite{Cortadella}
particularly satisfying.  In some cases, researchers have succeeded in
more exhaustive {\it ad hoc} approaches, such as \cite{Cox_2000}.
Other methods have focused on robustness against faults
\cite{Monnet_2006} or hierarchical design \cite{Butucaru_2007}.

Once a circuit has been synthesized, its behavior should be verified
against the original specification.  The most straightforward
simulation involves generating accurate timeseries of the
voltage or logic signals in the circuit.  Since there are manufacturing
variations, it is helpful to propagate ambiguous logic values during
switching transitions \cite{Chappell_1971}.  Most modern approaches
for verification generally involve symbolic event-level
simulation that is motivated by temporal logic \cite{Browne_1986}.
However, this approach usually suffers from a state explosion unless
the original specifications are taken into account
\cite{Clarke_1987}.  Using algebraic manipulations \cite{Ishiura_1989}
also cuts back on the computational load of such a simulation.  More
recently, verification tools have been unified into the hardware
description languages used for design \cite{Endecott_1998}, and
address hierarchical design workflows \cite{Vakilotojar_2000}.  We
refer the reader to surveys \cite{Meister_1993} and \cite{Baker_1994}
for more extensive treatment of asynchronous simulation.

In contrast to verification by temporal logic or timeseries
simulation, the approach taken in this article suggests that the
topological structure of a logic circuit plays an important role in
determining its behavior.  This appears to be related to the recent
approach in \cite{Meredith_2007}, but topological analysis of
electrical circuits is not new.  Indeed, algebraic topology can be
used to show that the usual formulation of electrical circuit laws
results in a solution for voltages and currents \cite{Roth_1955}.
Branin \cite{Branin_1966} showed how a topological approach can be
extended to address a wide class of network-related problems.
Moreover, Smale \cite{Smale_1972} showed that the differential
equations describing electrical network behavior can be derived from
homology theory.  Smale's results were subsequently extended for more
general circuit elements by Calvert \cite{Calvert_1997}.  This
dynamical viewpoint can also be understood using the topology of
manifolds \cite{Matsumato_1976}.

\subsection{Our approach}

In order to shorten the design cycle for asynchronous circuits, it is desirable to bridge the gap between static logic state computation and event-level simulation.  Ideally, such a technique would avoid both the level of detail and the computational burden of exhaustive simulation, while providing coarse semantic properties that static logic computation cannot address.  This article describes a way to encode slightly more information than the netlist (the gates and connections) and logic values on the wires.  

Specifically, we assume that consistent logic states may extend over portions of the circuit or over the entire circuit.  Consistent logic states that cannot be extended consistently over the entire circuit correspond to transient states: inconsistency occurs along wires where the logic value is changing.  {\it In this way, we are able to examine circuit behavior that involves unknown delays along wires in an implicit fashion: we never need to specify the delays as variables as in \cite{Ishiura_1989} nor do we make any assumption about delays remaining fixed during the operation of the circuit.}  Since this information is local in the usual topology induced on the directed graph describing circuit connections, the natural computational framework is that of {\it sheaf theory}.  From the outset, a direct application of sheaf theory to logic circuits results in significant computational difficulties since the natural sheaf is not one of abelian groups.

However, by lifting the logic values from binary values into an abstract vector space spanned by logic 0 and logic 1, we obtain new information from the sheaf (at the level of its global sections, rather than just in its relative cohomology) and computations become straightforward exercises in linear algebra.  This seemingly abstract trick corresponds to using one-hot signaling \cite{Davis_1997}, which is used to provide error detection in existing asynchronous interfaces.  Mathematically, using this encoding gives the resulting {\it switching sheaf} enough freedom to describe global logic states that are the superposition of two transitional states; essentially by capturing undefined signals and signal collisions.  Therefore, by examining the cohomology of switching sheaves, certain behaviors can be detected in addition to the static logic states.  The main result is that the first cohomology group of switching sheaves is generated by all the feedback loops that have the potential to latch or cause glitches.  On the other hand, we show that combinational logic circuits in which each input is used exactly once (and therefore cannot glitch) have trivial first cohomology.  Therefore, it appears that the first cohomology group contains certificates of truly asynchronous behavior.

As an aside, we note that without one-hot signalling, a sheaf theoretic approach to this problem could still proceed by looking for obstructions to extending local logic states.  This corresponds exactly to event-level simulation, which shows that simulation is substantially more computationally expensive than our approach.  We hold out some hope that a coarser obstruction theory exists (for switching sheaves, using one-hot signalling) that is more refined than the one presented in this article yet less exhaustive than a complete simulation.

\subsection{Outline of the paper}

In Section \ref{sheaf_sec} we give the basic definitions and highlight
the relevant results from sheaf theory.  Section
\ref{def_sec} describes our encoding of a logic circuit as a
switching sheaf.  In Section \ref{h1_sec}, we show how the cohomology
group of a switching sheaf captures the logic states that arise from
sustained feedback.  We give three examples of switching sheaves and
computation of their cohomology in Section \ref{example_sec},
culminating in a demonstration that the cohomology of a
switching sheaf carries more information than the list of logic
states.  Finally, the results are discussed in Section
\ref{discussion_sec}.

\section{Highlights from sheaf theory}
\label{sheaf_sec}

A sheaf is a mathematical tool for storing local information over a
domain.  It assigns some algebraic object, a vector space in our case,
to each open set, subject to certain compatibility conditions.  These
conditions are of two kinds: (1) those that pertain to restricting the
information from a larger to a smaller open set, and (2) those that
pertain to assembling information on small open sets into information
on larger ones.  What is of particular interest is the relationship of
the global information, which is valid over the entire graph, to the
topology of that graph.  This is captured by the cohomology of the
sheaf, in the way we summarize here.

\subsection{Elementary definitions for sheaves}
In this section, we follow the introduction to sheaves given in
Appendix 7 of \cite{Hubbard}, largely for its direct treatment of
sheaves over tame spaces.  For more a more general, and more
traditional approach, compare our discussion with \cite{Bredon}.

A {\it presheaf} $F$ is the assignment of a vector space $F(U)$ to each
open set $U$ and the assignment of a linear map $\rho_U^V: F(U) \to F(V)$ for
each inclusion $V \subseteq U$.  We call the map $\rho_U^V$ the {\it
  restriction map} from $U$ to $V$.  Elements of $F(U)$ are called
{\it sections} of $F$ {\it defined over} $U$.  

A {\it sheaf} $\mathcal{F}$ is a presheaf $F$ that satisfies the gluing axioms:
\begin{itemize}
\item (Monopresheaf) Suppose that $u \in F(U)$ and that $\{U_1,
  U_2,...\}$ is an open cover of $U$.  If $\rho_U^{U_i} u = 0$ for
  each $i$, then $u=0$ in $F(U)$.  Simply: sections that agree
  everywhere locally also agree globally.
\item (Conjunctivity) Suppose $u \in F(U)$ and $v \in F(V)$ are
  sections such that $\rho_U^{U\cap V} u = \rho_V^{U \cap V} v$.  Then
  there exists a $w \in F(U \cup V)$ such that $\rho_{U \cup V}^U w =
  u$ and $\rho_{U \cup V}^V w = v$.  In other words, sections that
  agree on the intersection of their domains can be ``glued together''
  into a section that is defined over the union of their domains of
  definition.
\end{itemize}

Standard examples of sheaves are 
\begin{itemize}
\item The collection of continous real-valued functions
  $C(X,\mathbb{R})$ over a topological space $X$.
\item The collection of locally constant functions, which essentially
  assigns a constant to each connected component of each open set.  
\end{itemize}
In contrast, the collection of {\it constant functions} does not form
a sheaf.  

There are six famous operations on sheaves that are important in the
general theory, but only one of them (cohomology)
play a role in this article.

\subsection{Cohomology}
We can recast the conjunctivity axiom as measuring the kernel of the
linear map $d:\mathcal{F}(U) \oplus \mathcal{F}(V) \to \mathcal{F}(U
\cap V)$ given by $d(x,y)=\rho_U^{U\cap V}x-\rho_V^{U\cap V}y$.
Indeed, all of the elements of the kernel of such a linear map
correspond to the agreement of sections on $U \cap V$.
On the other hand, the monopresheaf axiom indicates that the preimage
of zero under the map $d$ corresponds to the restriction of these
glued sections onto each of $U$ and $V$.  Indeed, any nonzero element
of the {\it image} of $d$ cannot be a section over $U \cup V$.  

These two points motivate a computational framework for working with
sheaves, called the \v{C}ech construction.  

Suppose $\mathcal{F}$ is a sheaf on $X$, and that $\mathcal{U}=\{U_1,U_2,...\}$ is a cover
of $X$.  We define the {\it \v{C}ech cochain spaces}
$\check{C}^k(\mathcal{U};\mathcal{F})$ to be the direct sum of
the spaces of sections over each $k$-wise intersection of elements in
$\mathcal{U}$.  That is 
$\check{C}^k(\mathcal{U};\mathcal{F})=\bigoplus \mathcal{F}(U_{i_1} \cap ... \cap U_{i_k})$.

We define a sequence of linear maps $d^k:\check{C}^k(\mathcal{U};\mathcal{F}) \to \check{C}^{k+1}(\mathcal{U};\mathcal{F})$
by 
\begin{equation*}
d^k(\alpha)(U_1,U_2,...,U_{k+1})=\sum_{i=0}^{k+1} (-1)^i \rho^{U_0
  \cap ... \hat{U}_i ... \cap U_{k+1}}_{U_0 \cap ... \cap U_{k+1}}
\alpha(U_0
  \cap ... \hat{U}_i ... \cap U_{k+1}),
\end{equation*}
where the hat means that an element is omitted from the list.
Note that these fit together into a sequence, called the {\it \v{C}ech
cochain complex}: 
\begin{equation*}
\begin{CD}
0 \to \check{C}^0(\mathcal{U};\mathcal{F}) @>d^0>> \check{C}^1(\mathcal{U};\mathcal{F}) @>d^1>> ...
\end{CD}
\end{equation*}
A standard computation shows that $d_k \circ d_{k-1} = 0$, so that we can define
the {\it $k$-th \v{C}ech cohomology space}
$\check{H}^k(\mathcal{U};\mathcal{F}) = \text{ker } d_k / \text{image } d_{k-1}.$

The $\check{H}^k$ apparently depend on the choice of cover
$\mathcal{U}$, but for {\it good} covers (much as in the Nerve Lemma),
this dependence vanishes. Leray's theorem for sheaves states that
$\check{H}^k(\mathcal{U};\mathcal{F})$ is the same for each good
cover.  So we write
$H^k(X;\mathcal{F})=\check{H}^k(\mathcal{U};\mathcal{F})$, the sheaf's
{\it cohomology} in the case that $\mathcal{U}$ is a good cover.

A little thought about good covers on graphs reveals two important facts:
\begin{itemize}
\item for a metric graph $X$, $H^k(X;\mathcal{F})=0$ for $k>1$, and 
\item $H^0(X;\mathcal{F})$ is isomorphic to the space of global sections $\mathcal{F}(X)$.  
\end{itemize}

By analogy with the Mayer-Vietoris sequence for homology, there is a
Mayer-Vietoris sequence for sheaf cohomology \cite{Bredon}.  Suppose
that $A,B$ are two open subspaces of a graph $X$ that cover $X$, and
that $F$ is a sheaf over $X$.  Then the following {\it Mayer-Vietoris
  sequence} is an exact sequence:
\begin{equation*}
\begin{CD}
...\to H^k(X;\mathcal{F})@>r>>H^k(A;\mathcal{F})\oplus H^k(B;\mathcal{F})@>d>>H^k(A \cap B;\mathcal{F})@>\delta>>H^{k+1}(X;\mathcal{F})\to...
\end{CD}
\end{equation*}
In this sequence, $r$ comes from restriction maps in the obvious way,
$d$ is the composition of restriction maps and a difference: $d(x,y)=
\rho_A^{A\cap B} x - \rho_B^{A\cap B} y$, and $\delta$ is the
connecting homomorphism.  Notation has been abused above
slightly: by $H^k(A;\mathcal{F})$ we mean the $k$-th cohomology of the sheaf $F$
restricted to subsets lying in $A$.

\section{Construction of a switching sheaf from a circuit}
\label{def_sec}

This section describes a way to associate a sheaf structure to a directed graph that encodes a logic circuit.  Each vertex represents a logic gate, where the in-degree represents the number of inputs.  Each edge of the graph corresponds to a 1-bit signal connecting the the input of one gate to the output of another.  We allow edges to be self-loops (connecting the input of a gate to an output of the same gate) and external connections.  As existing logic circuits contain finitely many gates, we assume that the underlying graph is finite, but not necessarily connected.

\subsection{Quiescent logic states, one-hot encoding, and categorification}
We begin with a brief description of the circuit model to be encoded.
As the sheaf structure will require logic functions to be {\it linear}
functions, we categorify them.  This is accomplished by the relatively
standard one-hot encoding of logical values.

Suppose that $X$ is a directed graph in which each vertex has finite
degree.  A {\it logic circuit} is the assignment of a function
$f_v:\mathbb{F}_2^{m(v)} \to \mathbb{F}_2^{n(v)}$ to each vertex $v$, 
 where $m(v)$ is the in-degree of $v$ and $n(v)$ is the out-degree of
 $v$.  We call $f_v$ the {\it logic gate} at $v$.

Given a logic circuit, a {\it quiescent logic state} (QLS) is an assignment $s:E \to
\mathbb{F}_2$ of a binary value to each edge, such that for each
vertex $v$, $f_v(s(e^+_1),s(e^+_2),...)=(s(e^-_1),s(^-_2),...)$ where
$\{e^+_i\}$ are the incoming edges at $v$ and $\{e^-_i\}$ are the
outgoing edges at $v$.

In this article, we examine {\it one-hot} encoding $T$ of binary values in
a logic circuit.  That is, we consider the function
$T:\mathbb{F}_2\to\mathbb{F}_2^2$ where 
\begin{eqnarray*}
T(0)&=&\begin{pmatrix}1\\0\end{pmatrix}\\
T(1)&=&\begin{pmatrix}0\\1\end{pmatrix}.  
\end{eqnarray*}

Applying this replacement to
    each occurance of $\mathbb{F}_2$ in the definition of a logic
    circuit and logic state results in a particular {\it
      categorification} of logic circuits.  Indeed, each of the logic
    gates $f_v$ become {\it linear} functions $T f_v$ between $\mathbb{F}_2$ vector
    spaces. 

Casual examination of the categorification procedure suggests that
very little has changed, except the algebraic structure has been
slightly enhanced.  However, two new things have occured:
\begin{itemize}
\item Problems of logic can now be addressed computationally using the
  framework of linear algebra.  This can result in gains in asymptotic
  computational complexity.  Rather than being forced to enumerate
  states, one may instead perform standard polynomial-time linear
  algebra (over the finite field $\mathbb{F}_2$).
\item It is possible to superpose two logic states, and thereby
  study certain kinds of transitions between logic states.  This is
  subtle and somewhat surprising: we have not explicitly described
  anything about time evolution of circuits, and indeed the usual way
  of examining the QLS of a logic circuit does not concern itself with
  time.  However, by permitting superposed states, we are able to
  study the circuit's response to both {\it simultaneously} and
  thereby discern the way that one might transition to the other. 
\end{itemize} 

\subsection{Switching sheaves}
Suppose that $X$ is a directed graph with the usual topology, let $\mathcal{U}=\{U_\alpha,V_\beta\}$ be a base for the topology of $X$ where each $U_\alpha$ is connected and contains exactly on vertex and each $V_\beta$ is contained in the interior of a single edge.
A {\it switching sheaf} $\mathcal{S}$ on $X$ is the sheafification of the following presheaf $S$, defined on $\mathcal{U}$:
\begin{itemize}
\item $S(U_\alpha)$ is the tensor product of copies $\mathbb{F}_2$, one for each incoming edge into the unique vertex contained in $U_\alpha$,
\item $S(V_\beta)=\mathbb{F}_2$,
\item if $V_\beta \subset U_\alpha$ and $V_\beta$ is contained in the
  $n$-th incoming edge, the the restriction map $S(U_\alpha) \to
  S(V_\beta)$ is the contraction onto the $n$-th factor of $\mathbb{F}_2$,
\item if $V_\beta \subset U_\alpha$ and $V_\beta$ is contained in the
  $n$-th outgoing edge, then there is a fixed $\mathbb{F}_2$-linear map
  $S(U_\alpha) \to S(V_\beta)$ depending only on the vertex $v$
  contained in $U_\alpha$ and $n$ (the outgoing edge).  This
  collection of maps $\{\phi_v\}$ for vertices $v$, satisfies $\phi_v=T f_v$.
\end{itemize}

\begin{figure}
\begin{center}
\includegraphics[width=3.5in]{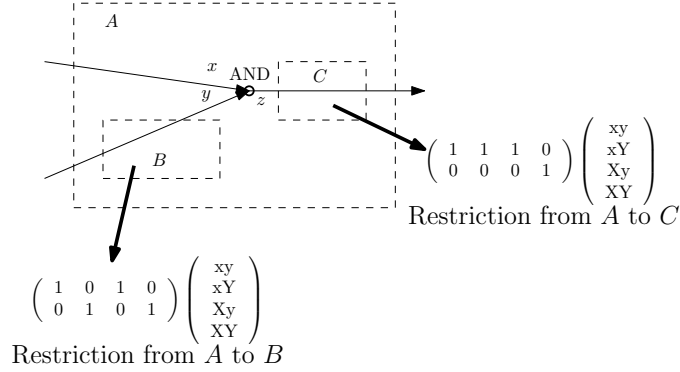}
\caption{An example of a switching sheaf (in this figure $xY$ denotes $\overline{x}\otimes y$)}
\label{initial_circ_fig}
\end{center}
\end{figure}

Figure \ref{initial_circ_fig} gives an example of a switching sheaf
over a graph with one vertex, which represents an AND gate.  Notice that the dimension of the sheaf over $B$ and $C$ is 2, while it has dimension 4 over $A$.  
 
When we treat \v{C}ech cohomology with respect to the cover $\mathcal{U}$, we will use the notation $\check{H}^k(X;\mathcal{S})$, rather than $\check{H}^k(\mathcal{U};\mathcal{S})$, to emphasize that this choice of cover is being used.

\begin{prop}
\label{qls_subset_h0_prop}
Suppose $\mathcal{S}$ is a switching sheaf over a logic circuit $X$.
Every QLS of this logic circuit lifts via $T$ to an element of
$H^0(X;\mathcal{S})$.  Conversely, every element of
$H^0(X;\mathcal{S})$ that restricts to
$\left(\begin{smallmatrix}1\\0\end{smallmatrix}\right)$ or
  $\left(\begin{smallmatrix}0\\1\end{smallmatrix}\right)$ on each edge is the image of a
    QLS through this lift.  Any such section is nonvanishing.
\end{prop}
\begin{proof}
Given a QLS $s$, it is clear that $\sigma = T \circ s$ defines a section of
$\mathcal{S}$ over the edges of $X$.  We only need to address the
value of this lifted section at the vertices.  The correct answer is
easy to obtain.  Suppose $v$ is a vertex with incoming edges
$\{e^+_1,e^+_2,...,e^+_k\}$.  Then the appropriate definition for $\sigma(v)$
is $(T \circ s)(e^+_1) \otimes (T \circ s)(e^+_2) \otimes ... \otimes (T
\circ s)(e^+_k)$.  The definition of $\phi_v$ as $T f_v$ ensures that
the lifts of each outgoing edge through $T$ agrees with our choice for
$\sigma(v)$.  Therefore, $\sigma$ lifts to an element of
$H^0(X;\mathcal{S})$, which we define as $Ts$.

Conversely, suppose we have a global section $\tau$ of $\mathcal{S}$
that restricts to $\left(\begin{smallmatrix}1\\0\end{smallmatrix}\right)$ or
  $\left(\begin{smallmatrix}0\\1\end{smallmatrix}\right)$ on each edge.  Suppose that $v$
    is a vertex and that $U$ is a contractible open set containing $v$
    only.  Since the map $\mathcal{S}(U) \to \mathcal{S}{V^+_i}$ for
    each incoming edge $e_i$ is a contraction onto the $i$-th factor,
    we have that $\tau(v) = (Ta_1)\otimes (Ta_2) \otimes \cdots \otimes
    (Ta_n)$ where $(Ta_i)$ is the value of $\tau$ on the $i$-th
    incoming edge.  Clearly, this is well-defined since the image of
    $T$ on $\mathbb{F}_2$ is the two-element set
    $\left\{\left(\begin{smallmatrix}1\\0\end{smallmatrix}\right),
  \left(\begin{smallmatrix}0\\1\end{smallmatrix}\right)\right\}$. 

If $V$ contains the interiors of all outgoing edges for $v$, then by
the definition of the switching sheaf $\mathcal{S}$, $\mathcal{S}(U)
\to \mathcal{S}(U \cap V)$ is a linear map
\begin{eqnarray*}
\phi_v(\tau(v))&=&\phi_v((Ta_1)\otimes (Ta_2) \otimes \cdots \otimes
(Ta_n))\\
&=&(Tf_v)((Ta_1)\otimes (Ta_2) \otimes \cdots \otimes
(Ta_n))\\
&=&T(f_v(a_1,a_2,... a_n)),\\
\end{eqnarray*}
which indicates that $\tau$ is the image of some QLS, whose incoming
edges at $v$ have values $a_1,a_2,... a_n$.  This computation makes use
of the commutative diagram
\begin{equation*}
\begin{CD}
\mathbb{F}_2^{2^{2n}} @>Tf_v=\phi_v>> \mathbb{F}_2^{2^{2m}}\\
@A T AA @A T AA\\
\mathbb{F}_2^{2n} @>f_v>> \mathbb{F}_2^{2m}\\
\end{CD}
\end{equation*}

Such a section of $\mathcal{S}$ is nonvanishing: for any QLS $s$, the
function $(T \circ s)$ is clearly nonvanishing on the edges.  At
vertices, the lift takes values that are the tensor product of the
incoming edge values, which are all nonzero.
\end{proof}

\section{The content of the cohomology of a switching sheaf}
\label{h1_sec}

In this section, we use the Mayer-Vietoris sequence to examine how
switching sheaf cohomology changes as a circuit is progressively
assembled.  In this way, we describe an incremental method for
computing switching sheaf cohomology that mimics the way a prototype
circuit could be ``soldered'' together.  The effect of adding an
unconnected gate is straightforward, but adding a single connection
wire reveals the meaning of $H^1$: nontrivial elements of $H^1$
correspond to sustained feedback states. Their presence is therefore
an indication of possible latching (stable feedback) or glitches
(unstable feedback, usually caused by race conditions).  Currently, we
do not know how to use the switching sheaves to discriminate
between the two kinds of feedback, as the formalism apparently corresponds to an
unbounded wire delay model.

We first consider the effect of adding a new gate $G$ to a circuit
$A$, but not connecting the two.  We therefore consider the switching
sheaf $\mathcal{S}$ on $X=A \sqcup G$, where $G$ is a single vertex.
The Mayer-Vietoris sequence in this case consists only of the
isomorphism $H^k(X;\mathcal{S}) \cong H^k(A;\mathcal{S}) \oplus
H^k(G;\mathcal{S})$ for all $k$.  However, we note immediately that
$H^k(G;\mathcal{S})$ is trivial for $k>0$ since the covering dimension
of $G$ is zero.  Thus, $H^1$ is unchanged by adding an unconnected
logic gate to a circuit, and the the dimension of $H^0$ increases by
$2^{\text{\# inputs of } G}$.  

\begin{figure}
\begin{center}
\includegraphics[width=2in]{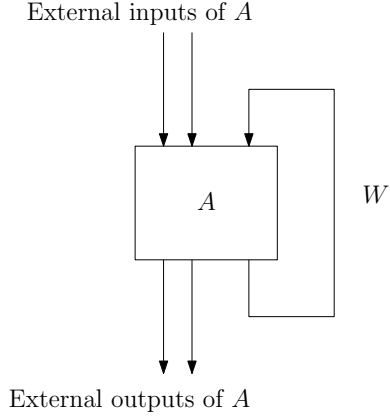}
\caption{Circuit $A$ with a feedback wire $W$ attached}
\label{mv_circ_fig}
\end{center}
\end{figure}

In order to explain the effect of attaching a wire $W$ to an existing
circuit $A$ (see Figure \ref{mv_circ_fig}), we construct the
Mayer-Vietoris sequence for $X=A\cup W$ and a switching sheaf on
$\mathcal{S}$.  In order to ensure the correct interpretation, we
assume $W$ is a connected subset of an edge and $A$ is homotopy
equivalent to $X-W$.  The Mayer-Vietoris sequence in this case is (we
suppress the sheaves from the notation)
\begin{equation*}
\begin{CD}
0 \to H^0(X) \to H^0(A)\oplus \mathbb{F}_2^2 @>\Delta>> \mathbb{F}_2^4
\to H^1(A\cup W) \to H^1(A) \to 0.
\end{CD}
\end{equation*}
Note that exactness requires that $\dim H^1(A\cup W;\mathcal{S})
\ge \dim H^1(A; \mathcal{S})$.  Observe that the difference map takes the form
\begin{equation*}
\Delta=\begin{pmatrix}
P_{2 \times k} & I_{2\times 2}\\
Q_{2 \times k} & I_{2\times 2}\\
\end{pmatrix},
\end{equation*}
where $I_{2 \times k}$ is a 2 by 2 identity matrix, $P_{2 \times k}$ and $Q_{2 \times k}$ are 2 by $k$ matrices, and $k$ is the dimension of $H^0(A;\mathcal{S})$.  The matrix $P_{2 \times k}$ represents the restriction of sections over $A$ to the output of the wire $W$, or equivalently to the particular input of $A$ where the wire attaches.  In much the same way, $Q_{2 \times k}$ is the restriction from the sections of $A$ to the particular output of $A$ that is attached to the wire.  Observe that a pair of row reductions on $\Delta$ results in the matrix
\begin{equation*}
\begin{pmatrix}
P_{2 \times k} & I_{2\times 2}\\
Q_{2 \times k}-P_{2 \times k} & 0_{2\times 2}\\
\end{pmatrix},
\end{equation*}
which has rank 2, 3, or 4.  The rank of $\Delta$ depends how much the wire participates in the feedback of signals, so we assign names to the three possibilities:
\begin{itemize}
\item rank $\Delta=2$: {\it complete feedback}, in which $Q_{2 \times k}=P_{2 \times k}$.  This occurs when the input and output of $A$ that the wire connects always agree.
\item rank $\Delta=3$: {\it partial feedback}.
\item rank $\Delta=4$: {\it no feedback}.  This case occurs especially when the wire $W$ connects two disconnected components of $A$, but more generally when the input and output connected by $W$ are completely independent.
\end{itemize}
Therefore,
\begin{equation*}
\dim H^0(X;\mathcal{S})=\dim H^0(A;\mathcal{S})-\begin{cases}
0& \text{if complete feeback}\\
1& \text{if partial feedback}\\
2& \text{otherwise}\\
\end{cases}
\end{equation*}
and
\begin{eqnarray*}
\dim H^1(X;\mathcal{S})&=&\dim H^1(A;\mathcal{S})+4-\rank \Delta \text{ (by exactness)}\\
&=&\dim H^1(A;\mathcal{S})+\begin{cases}
2& \text{if complete feeback}\\
1& \text{if partial feeback}\\
0& \text{otherwise}\\
\end{cases}\\
\end{eqnarray*}
The effect of attaching $W$ is best described by the following slogan: 
\begin{itemize}
\item Attaching a wire that does not participate in feedback suppresses logic states and leaves $H^1$ unchanged.
\item Attaching a wire that participates in feedback leaves logic states unchanged and adds to the dimension of $H^1$.
\end{itemize}

\section{Examples of switching sheaves and their cohomology}
\label{example_sec}

In this section, we exhibit the cohomology of switching sheaves and its interpretation by way of  three illustrative examples: combinational circuits with and without shared inputs and an RS flip-flop.  These examples indicate that $H^0$ of a switching sheaf contains at least as many elements as the set of QLS.  Additionally, as was shown in Section \ref{h1_sec}, $H^1$ of a switching sheaf captures information about the presence of feedback or race conditions.  We give two explicit examples of this fact.

\subsection{Combinational circuits without shared inputs}

Let us consider the case of a switching sheaf $\mathcal{S}$ on a
connected, directed tree $X$.  (The choice of directions on the edges
of $X$ does not effect the cohomology of $\mathcal{S}$.)  This
represents the situation in which each external input is used at most
once in the production of each external output.  In this case, $\{X\}$
by itself is a good cover, so we conclude that $H^1(X;\mathcal{S})$ is
trivial.  Observe that the combinatorial Euler characteristic of $X$
is 1 by the same reasoning, so that the number of vertices of $X$ is 1
more than the number of internal edges.  Thus, if there are $n$ vertices with in-degrees $\{m_1, ... m_n\}$,
\begin{equation*}
\dim H^0(X;\mathcal{S})=\dim \check{C}^0- \dim \check{C}^1=\sum_{i=1}^n 2^{m_i} - 2(n-1).
\end{equation*}

\begin{figure}
\begin{center}
\includegraphics[width=3in]{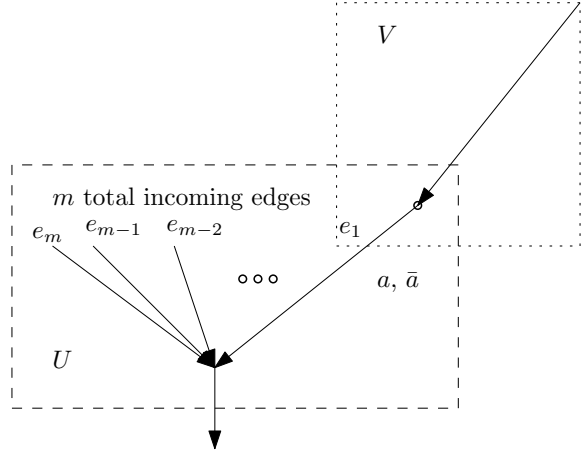}
\caption{A single logic gate with an extended input edge}
\label{m_input_fig}
\end{center}
\end{figure}

Looking at a basis for $H^0(X;\mathcal{S})$ is instructive, so
consider the logic circuit shown in Figure \ref{m_input_fig}.
This circuit consists of a single $m$-input logic gate.  One of the
input edges (labeled with signals $a$ and $\bar{a}$) is extended to
include a single 1-input buffer gate (identity function).  We compute
the sheaf cohomology and a basis for this cohomology using a \v{C}ech
complex.  This complex has the form
\begin{equation*}
\begin{CD}
0 \to \underbrace{\mathbb{F}_2^2 \otimes \cdots \otimes
  \mathbb{F}_2^2}_{\mbox{$m$ factors}} @>d^0>> \mathbb{F}_2^2
\to 0.
\end{CD}
\end{equation*}
The matrix form of $d^0$ is
\begin{equation*}
d^0=\begin{pmatrix}
1 & \cdots \text{total of } 2^{m-1} \text{ ones} \cdots & 1 & 0 & \cdots & 0 & 1 & 0\\
0 & \cdots & 0 & 1 & \cdots \text{total of } 2^{m-1} \text{ ones} \cdots & 1 & 0 & 1\\
\end{pmatrix}
\end{equation*}
which evidently has full rank.  Hence, the dimension of
$H^1(X;\mathcal{S})$ is zero.  The dimension of
$H^0(X;\mathcal{S})$ is $2^m$, which is the same as the number of QLS
for the logic circuit.  We would therefore expect that
$\check{H}^0(X;\mathcal{S})$ is spanned by images under $T$ of QLS, and this
is the case.  A basis is
\begin{center}
\begin{tabular}{c}
$\overline{a}+\overline{e_1} \otimes \overline{e_2} \otimes \cdots
\otimes \overline{e_m}$\\
$\overline{a}+\overline{e_1} \otimes e_2 \otimes \cdots
\otimes \overline{e_m}$\\
...\\
$\overline{a}+\overline{e_1} \otimes e_2 \otimes \cdots \otimes e_m$\\
$a+e_1 \otimes \overline{e_2} \otimes \cdots
\otimes \overline{e_m}$\\
$a+e_1 \otimes e_2 \otimes \cdots
\otimes \overline{e_m}$\\
...\\
$a+e_1 \otimes e_2 \otimes \cdots \otimes e_m$\\
\end{tabular}
\end{center}
where $a$ is supported on $V$, and the other term is
supported on $U$.  Notice in particular that all sections are
supported over the entirety of $X$, and all restrict to
$\left(\begin{smallmatrix}0\\1\end{smallmatrix}\right)$ or
  $\left(\begin{smallmatrix}1\\0\end{smallmatrix}\right)$ on edges.  Hence, this basis
    consists of images of QLS.

\begin{figure}
\begin{center}
\includegraphics[width=3in]{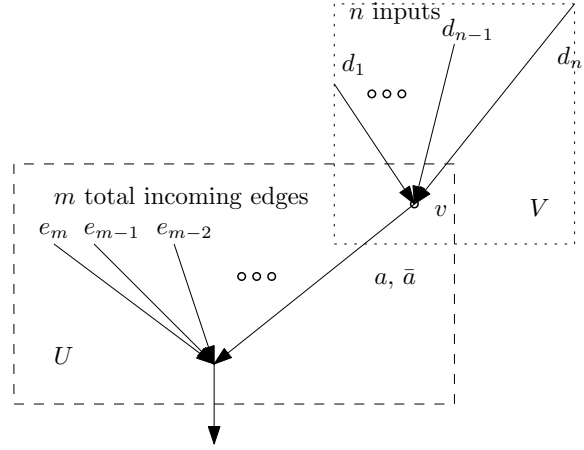}
\caption{Two logic gates composed}
\label{m_n_input_fig}
\end{center}
\end{figure}

Given a situation like that shown for the graph $Y$ in Figure
\ref{m_n_input_fig}, we observe that the number of QLS present is
$2^{n+m-1}$.  However, the dimension of $H^0(Y;\mathcal{S})$ differs
from this number.  We construct the \v{C}ech coboundary map $d^0$ in matrix form
\begin{equation*}
d^0=\left(\begin{array}{cccccc|c}
1&\cdots&1&0&\cdots&0&\overline{f_v}\\
0&\cdots&0&1&\cdots&1&f_v
\end{array}\right),
\end{equation*}
where by $f_v$ we mean a $1\times 2^n$ submatrix with zeros in the
entries corresponding to gate $v$ taking output value 0.  The
coboundary map is evidently of full rank, so that $H^0(Y;\mathcal{S})$
has dimension $2^n+2^m-2$, and $H^1(Y;\mathcal{S})$ is trivial.  

Suppose that $f_v$ has $k$ nonzero entries.  We note that
$\check{H}^0(x;\mathcal{S})$ has a basis of lifted QLS.  There are $k+2^{n-1}-1$ basis
elements of the form (in particular, the first term is where $f_v$ is
nonzero and $e_1$ participates in the second term)
\begin{equation*}
d_1 \otimes \cdots \otimes d_n + e_1 \otimes \cdots \otimes e_m.
\end{equation*}
There are additionally $2^m-k+2^{n-1}-1$ elements of the form (in which
$\overline{e_1}$ participates in the second term)
\begin{equation*}
d_1 \otimes \cdots \otimes d_n + \overline{e_1} \otimes \cdots \otimes
e_m.
\end{equation*}

This proves the obvious fact that if no inputs are shared in a
combinational circuit, then the entire circuit has no interesting
asynchronous behavior.  It should therefore be possible to prove the
following conjecture, though we have not yet succeeded.

\begin{conj}
\label{dag_conj}
If $\mathcal{S}$ is a switching sheaf over a directed tree $X$,
then $H^0(X;\mathcal{S})$ has a basis that consists of lifted QLSs.
\end{conj}

This means that any section over $X$ that vanishes anywhere must be
the linear superposition of two or more QLS, and therefore describes
uncertainty or transient states.  

\subsection{Combinational circuits with shared inputs}

\begin{figure}
\begin{center}
\includegraphics[width=3in]{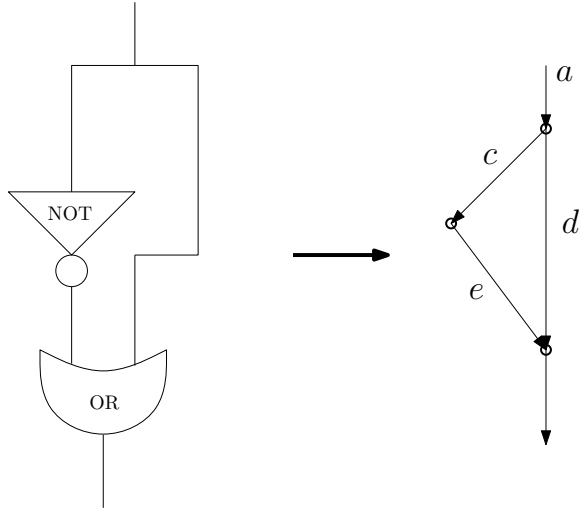}
\caption{A combinational logic circuit with a shared input signal}
\label{glitch_circ_fig}
\end{center}
\end{figure}

The circuit shown in Figure \ref{glitch_circ_fig} does not satisfy the
hypotheses of Conjecture \ref{dag_conj}.  In particular, it contains
two separate signal paths for the input $a$.  It should be clear that
as a logic circuit, this has two QLS: one for each binary input value.
However, assuming that there is some delay in the circuit, the signal
labeled $e$ will be delayed from the ideal signal $\overline{a}$.
This means that there is some time-sensitivity in the circuit, and it
can therefore produce glitches (narrow pulses on its output) when the
input is changed.

If we consider a switching sheaf over the logic circuit, we obtain a
\v{C}ech coboundary matrix that has the form
\begin{equation*}
d^0=\left(
\begin{array}{cc|cc|cccc}
-1& 0& 1& 0&  0& 0& 0& 0\\
0& -1& 0& 1& 0& 0& 0& 0\\
\hline
-1& 0& 0 &0&   1& 1& 0& 0\\
0& -1& 0& 0&   0& 0& 1& 1\\
0& 0&  0& -1&  1& 0 &1& 0\\
0& 0& -1& 0 &  0& 1& 0& 1\\
\end{array}
\right),
\end{equation*}
where minus signs have been added for convenience.  Row reduction of
this matrix reveals a basis of sections supported over the entire
graph:
\begin{center}
\begin{tabular}{c}
$\overline{a}+\overline{c}+\overline{d}\otimes e$\\
$a+c+d\otimes \overline{e}$\\
$a+\overline{a}+c+\overline{c}+d\otimes e+\overline{d}\otimes\overline{e}$\\
\end{tabular}
\end{center}
It is apparent that the first two basis elements are lifts of the
QLS.  However, the second is clearly neither a lift of a QLS, nor a
linear combination of them.  Indeed, it indicates that ambiguity in
the input logic value (such as occurs during a transition) causes
ambiguity throughout the rest of the circuit.  It is therefore an {\it
  algebraic} indication of the presence of time-sensitivity of the
circuit. 

In addition to the presence of the additional basis element for
$\check{H}^0(Y;\mathcal{S})$, there is another indication of additional
information.  $\check{H}^1(Y;\mathcal{S})$ is nontrivial in the case of this
logic circuit, and is generated by
$\overline{c}+c+\overline{d}+d+\overline{e}+e$, which indicates that
the source of the time-sensitivity is the two separate signal paths
for the input.

This calculation proves the following 

\begin{thm}
The cohomology of a switching sheaf over a logic circuit contains
different information than the set of its quiescent logic states.
\end{thm}

\subsection{An R-S flip-flop}

\begin{figure}
\begin{center}
\includegraphics[width=1.5in]{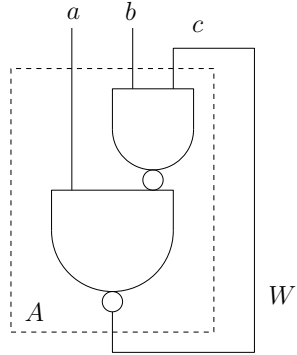}
\caption{An R-S flip-flop circuit}
\label{rsff_fig}
\end{center}
\end{figure}

There are other switching sheaf structures that can be constructed over a graph with one (undirected) loop.  While glitches are one kind of time sensitive behavior, another is the latching of a transient input.  We therefore give a classic example of a circuit that exhibits latching, the R-S flip-flop.

Consider the circuit $X$ shown in Figure \ref{rsff_fig}, which we split into two pieces: a combinational circuit $A$ with a 3-input gate, and a feedback wire $W$.  The QLS for this circuit \cite{Shannon_1940} are summarized in the following table:

\begin{center}
\begin{tabular}{ccc|c|l}
$a$&$b$&$c$&$q$&Description\\
\hline
0&0&1&1&Danger\\
0&1&1&1&Set\\
1&0&0&0&Reset\\
1&1&0&0&Hold zero\\
1&1&1&1&Hold one\\
\end{tabular}
\end{center}

Looking at the difference map $\Delta$ for the Mayer-Vietoris sequence, we note that
\begin{equation*}
P=\begin{pmatrix}
1&0&1&0 & 1&0&1&0\\
0&1&0&1 & 0&1&0&1\\
\end{pmatrix}
\end{equation*}
and
\begin{equation*}
Q=\begin{pmatrix}
0&0&0&0 & 1&1&1&0\\
1&1&1&1 & 0&0&0&1\\
\end{pmatrix}.
\end{equation*}  
The resulting matrix for $\Delta$ has rank 3, so that $H^0(X;\mathcal{S})$ has dimension 7 and $H^1(X;\mathcal{S})$ has dimension 1.  Here is a basis for $\check{H}^0(X;\mathcal{S})$:
\begin{center}
\begin{tabular}{c|c}
Element of $\check{H}^0(X;\mathcal{S})$&Description\\
\hline
$\overline{a}\otimes\overline{b}\otimes c$&Danger\\
$\overline{a}\otimes b \otimes c$&Set\\
$a\otimes \overline{b}\otimes \overline{c}$&Reset\\
$a\otimes b\otimes\overline{c}$&Hold zero\\
$a\otimes b\otimes c$&Hold one\\
$\overline{a}\otimes\overline{b}\otimes\overline{c}+a\otimes\overline{b}\otimes c$&Transition between Danger and Reset\\
$\overline{a}\otimes\overline{b}\otimes\overline{c}+\overline{a}\otimes b\otimes\overline{c}$&Transition between Danger and Set\\
\end{tabular}
\end{center}
Of most interest are the last two basis elements.  These are linear
combinations of two terms, neither of which is a lift of a QLS.  The
most suggestive interpretation is that they imply an uncertainty when
exiting the Danger state.  As the inputs $a$ and $b$ transition from
both logic 0 to both logic 1, there is a race condition.  Only one of
them transitions first, so there is a brief transition into
the Set or Reset states before entering a Hold state.  If we add the
last two basis elements, we obtain $a\otimes\overline{b}\otimes
c+\overline{a}\otimes b\otimes\overline{c}$ which indicates that an
uncertainty about which of $a$ or $b$ transitions has occured results
in uncertainty in the signal $c$.

\section{Discussion}
\label{discussion_sec}

The cohomology of switching sheaves is a new source of information
about the behavior of logic circuits, especially those circuits that
are asynchronous.  Especially, the presence of nontrivial elements of
$H^1$ indicates that a circuit has feedback or a race condition.  This
is a somewhat coarse descriptor of circuit behavior, as should be
expected from such a global topological invariant as cohomology.
However, there remain important questions regarding details at finer
timescales.  In particular, can the cohomology of switching sheaves
discriminate between glitches and latching?  If $H^1$ is trivial,
$H^0$ does not contain the same information as the logic states.
Indeed, a {\it basis} for $H^0$ often contains less information.  (The
set of QLS is contained in $H^0$, {\it considered as a set.}  See
Proposition \ref{qls_subset_h0_prop}.)  A sharper
connection to one of the popular semantic models of asynchronous logic
will likely be essential in answering these questions.

\bibliographystyle{plain}
\bibliography{circuits_bib}

\end{document}